\documentclass[12pt]{article}

\begin{document}


\def\IR{\mathchoice{ \hbox{${\rm I}\!{\rm R}$} }
                   { \hbox{${\rm I}\!{\rm R}$} }
                   { \hbox{$ \scriptstyle  {\rm I}\!{\rm R}$} }
                   { \hbox{$ \scriptscriptstyle  {\rm I}\!{\rm R}$}}}

\def\IN{\mathchoice{ \hbox{${\rm I}\!{\rm N}$} }
                   { \hbox{${\rm I}\!{\rm N}$} }
                   { \hbox{$ \scriptstyle  {\rm I}\!{\rm N}$} }
                   { \hbox{$ \scriptscriptstyle  {\rm I}\!{\rm N}$}}}

\def\IC{\mathchoice{ \hbox{${\rm l}\!\!\!{\rm C}$} }
                   { \hbox{${\rm l}\!\!\!{\rm C}$} }
                   { \hbox{$ \scriptstyle  {\rm l}\!\!\!\!\:{\rm C}$} }
                   { \hbox{$ \scriptscriptstyle {\rm l}\!\!\!\!\:{\rm C}$} } }

\def\kasten#1{\mathop{\mkern0.5\thinmuskip
                      \vbox{\hrule
                            \hbox{\vrule
                                  \hskip#1
                                  \vrule height#1 width 0pt
                                  \vrule}%
                            \hrule}%
                      \mkern0.5\thinmuskip}}

\def\Kasten{\mathchoice{\kasten{8pt}}%
                       {\kasten{6pt}}%
                       {\kasten{4pt}}%
                       {\kasten{3pt}}}%


\renewcommand{\labelenumi}{{\rm(\roman{enumi})}}

\newcounter{mathe}[section]
\newenvironment{mathe}[1]
{\par\bigskip\refstepcounter{mathe}\noindent\textbf{\arabic{mathe}
 #1.}\mdseries\ }
{\par\bigskip\mdseries}


\hyphenation{ana-logue ana-lo-gous ana-lo-gous-ly ana-ly-sis}

\title{Quantum Spacetime without Observers: Ontological Clarity and
the Conceptual Foundations of Quantum Gravity\footnote{To appear in {\it Physics Meets Philosophy at the Planck Scale,} edited by
C.\,Callender and N.\,Huggett (Cambridge University Press).}}
\author{Sheldon Goldstein \\
Department of Mathematics,
Rutgers University,\\ New Brunswick, NJ 08903, USA \and 
Stefan Teufel \\
Mathematisches Institut der Universit\"{a}t 
M\"{u}nchen,\\
Theresienstra{\ss}e 39,
80333 M\"{u}nchen, Germany}
\maketitle

\begin{abstract}
We explore the possibility of a Bohmian approach to the problem of finding
a quantum theory incorporating gravitational phenomena. The major {\it
conceptual\/} problems of canonical quantum gravity are the problem of time
and the problem of diffeomorphism invariant observables. We find that these
problems are artifacts of the subjectivity and vagueness inherent in the
framework of orthodox quantum theory. When we insist upon ontological
clarity---the distinguishing characteristic of a Bohmian approach---these
conceptual problems vanish. We shall also discuss the implications of a
Bohmian perspective for the significance of the wave function, concluding
with unbridled speculation as to why the universe should be governed by
laws so apparently bizarre as those of quantum mechanics.

\end{abstract}

\newpage

\section{Introduction}

{\footnotesize
\begin{quote}
The term ``3-geometry'' makes sense as well in quantum geometrodynamics as
in classical theory.  So does superspace. But space-time does not. Give a
3-geometry, and give its time rate of change. That is enough, under typical
circumstances to fix the whole time-evolution of the geometry; enough in
other words, to determine the entire four-dimensional space-time geometry,
provided one is considering the problem in the context of classical
physics. In the real world of quantum physics, however, one cannot give
both a dynamic variable and its time-rate of change. The principle of
complementarity forbids. Given the precise 3-geometry at one instant, one
cannot also know at that instant the time-rate of change of the
3-geometry. \ldots The uncertainty principle thus deprives one of any way
whatsoever to predict, or even to give meaning to, ``the deterministic
classical history of space evolving in time.'' {\it No prediction of
spacetime, therefore no meaning for spacetime,\/} is the verdict of the
quantum principle. \\ \hfill Misner, Thorne, Wheeler 1973

\end{quote}}

One of the few propositions about quantum gravity that most physicists in
the field would agree upon, that our notion of space-time must, at best, be
altered considerably in any theory conjoining the basic principles of
quantum mechanics with those of general relativity, will be questioned in
this article. We will argue, in fact, that most, if not all, of the
conceptual problems in quantum gravity arise from the sort of thinking on
display in the preceding quotation.

It is also widely agreed, almost 40 years after the first attempts to
quantize general relativity, that there is still no single set of ideas on
how to proceed, and certainly no physical theory successfully concluding
this program. Rather, there are a great variety of approaches to quantum
gravity; for a detailed overview, see, e.g., Rovelli  \cite{Rovelli}. While the
different approaches to quantum gravity often have little in common, they
all are intended ultimately to provide us with a consistent quantum theory
agreeing in its predictions with general relativity in the appropriate
physical domain.  Although we will focus here on the conceptual problems
faced by those approaches which amount to a canonical quantization of
classical general relativity, the main lessons will apply to most of the
other approaches as well.

This is because, as we shall argue, many of these difficulties arise from
the subjectivity and the ontological vagueness inherent in the very
framework of orthodox quantum theory, a framework taken for granted by
almost all approaches to quantum gravity. We shall sketch how most, and
perhaps all, of the conceptual problems of canonical quantum gravity vanish
if we insist upon formulating our cosmological theories in such a manner
that it is reasonably clear what they are about---if we insist, that is, upon
ontological clarity---and, at the same time, avoid any reference to such
vague notions as measurement, observers, and observables.

The earliest approach, canonical quantum gravity, amounts to quantizing
general relativity according to the usual rules of canonical
quantization. However, to apply canonical quantization to general
relativity, the latter must first be cast into canonical form. Since the
quantization of the standard canonical formulation of general relativity,
the Arnowitt, Deser, Misner formulation \cite{ADM}, has led to severe
conceptual and technical difficulties, nonstandard choices of canonical
variables, such as in the Ashtekar formulation \cite{Ashtekar} and in loop
quantum gravity \cite{loop}, have been used as starting points for
quantization. While some of the technical problems have been resolved by
these new ideas, the basic conceptual problems have not been addressed.

After the great empirical success of the standard model in particle
physics, the hope arose that the gravitational interaction could also be
incorporated in a similar model. The search for such a unified theory led
to string theory, which apparently reproduces not only the standard model
but also general relativity in a suitable low energy limit. However, since
string theory is, after all, a quantum theory, it retains all the
conceptual difficulties of quantum theory, and our criticisms and
conclusions pertaining to quantum theory in general, in Sections 3 and 4 of
this article, will apply to it as well.  Nonetheless, our focus, again,
will be on the canonical approaches, restricted for simplicity to pure
gravity, ignoring matter.

This article is organized as follows: In Section 2 we will sketch the
fundamental conceptual problems faced by most approaches to quantum
gravity. The seemingly unrelated problems in the foundations of orthodox
quantum theory will be touched upon in Section 3. Approaches to the
resolution of these problems based upon the demand for ontological clarity
will be discussed in Section 4, where we will focus on the simplest such
approach, the de Broglie-Bohm theory or Bohmian mechanics. Our central
point will be made in Section 5, where we indicate how the conceptual
problems of canonical quantum gravity disappear when the main insights of
the Bohmian approach to quantum theory are applied.  

Finally, in Section 6, we will discuss how the status and significance of
the wave function, in Bohmian mechanics as well as in orthodox quantum
theory, is radically altered when we adopt a universal perspective. This
altered status of the wave function, together with the very stringent
symmetry demands so central to general relativity, suggests the
possibility---though by no means the inevitability---of finding an answer to
the question, Why should the universe be governed by laws so apparently
peculiar as those of quantum mechanics?

\section{The conceptual problems of quantum gravity}

In the canonical approach to quantum gravity one must first reformulate
general relativity as a Hamiltonian dynamical system. This was done by ADM
\cite{ADM}, using the 3-metric $g_{ij}(x^a)$ on a space-like hypersurface
$\Sigma$ as the configurational variable and the extrinsic curvature of the
hypersurface as its conjugate momentum $\pi^{ij}(x^a)$\footnote{Actually
the extrinsic curvature is given by $K_{ij}= G_{ijab}\pi^{ab}$ where
$G_{ijab}$ is the so called supermetric, which is itself a function of
$g_{ij}$. This distinction is, however, not relevant to our discussion.}.
The real time parameter of usual Hamiltonian systems is replaced by a
``multi-fingered time'' corresponding to arbitrary deformations $d\Sigma$
of the hypersurface. These deformations are split into two groups: those
changing only the three dimensional coordinate system $x^a$ on the
hypersurface (with which, as part of what is meant by the hypersurface, it
is assumed to be equipped) and deformations of the hypersurface along its
normal vector field. While the changes of the canonical variables under
both kind of deformations are generated by Hamiltonian functions on phase
space, $H_i(g,\pi)$ for spatial diffeomorphisms and $H(g,\pi)$ for normal
deformations, their changes under pure coordinate transformations on the
hypersurfaces are dictated by their geometrical meaning. The dynamics of
the theory is therefore determined by the Hamiltonian functions $H(g,\pi)$
generating changes under normal deformations of the hypersurface.

Denote by $N(x^a)$ the freely specifiable
lapse function that determines  how far, in terms of proper length,
one moves the space-like hypersurface at the point $x=(x^a)$ along its normal
vector: This distance is $N(x^a)d\tau,$ where $\tau$ is a parameter
labeling the successive hypersurfaces arrived at under the deformation
(and defined by this equation).  The infinitesimal changes of the canonical
variables are then generated by the Hamiltonian $H_N$ associated with $N$
(an integral over $\Sigma$ of the product of $N$ with a Hamiltonian density
$H(g,\pi;x^a)$):

\begin{eqnarray} 
d g_{ij}(x^a) & = & \frac{\delta H_N (g,\pi)}{\delta
\pi^{ij}(x^a)}  d \tau \nonumber\\
d \pi^{ij}(x^a) & = & - \frac{\delta H_N (g,\pi)}{\delta
g_{ij}(x^a)} d \tau \, . \label{Hamil}
\end{eqnarray}
In what follows we shall denote by $H(g,\pi)$ the collection $\{H_N
(g,\pi)\}$ of all such Hamiltonians (or, what comes pretty much to the same
thing, the collection $\{H(g,\pi;x)\}$ for all points $x\in \Sigma$) and
similarly for $H_i$. 

It is important to stress that the theory can be formulated completely in
terms of geometrical objects on a three dimensional manifold, with no a
priori need to refer to its embedding into a space-time. A solution of
(\ref{Hamil}) is a family of 3-metrics $g(\tau)$ that can be glued together
to build up a 4-metric using the lapse function $N$ (to determine the
transverse geometry).  In this way the space-time metric emerges
dynamically when one evolves the canonical variables with respect to
multi-fingered time.

However, the initial canonical data cannot be chosen arbitrarily, but must
obey certain constraints: Only for initial conditions that lie in the
submanifold of phase space on which $H_i(g,\pi)$ and $H(g,\pi)$ vanish do
the solutions (space-time metrics $g_{\mu\nu}(x^\mu)$) also satisfy
Einstein's equations. In fact, away from this so called constraint manifold
the theory is not even well defined, at least not as a theory involving a
multi-fingered time, since the solutions would depend on the special way we
choose to evolve the space-like hypersurface, i.e., on the choice of
$N(x^a)$, to build up space-time. Of course, a theory based on a single
choice, for example $N(x^a)=1$, would be well defined, at least formally.
 
By the same token, the invariance of the theory under space-time
 diffeomorphisms is no longer so obvious as in the formulation in terms of
 Einstein's equations: In the ADM formulation 4-diffeomorphism invariance
 amounts to the requirement that one ends up with the same space-time, up
 to coordinate transformations, regardless of which path in multi-fingered
 time is followed, i.e., which lapse function $N$, or $\tau$-dependent
 sequence of lapse functions $N(\tau)$, is used. This says that for the
 space-time built up from any particular choice of multi-fingered time, the
 dynamical equations (\ref{Hamil}) will be satisfied for {\it any\/}
 foliation of the resulting space-time into space-like
 hypersurfaces---using in (\ref{Hamil}) the lapse function $N(\tau)$
 associated with that foliation---and not just for the foliation associated
 with that particular choice.

Formally, it is now straightforward to quantize this constrained Hamiltonian
theory using Dirac's rules for the quantization of constrained systems
\cite{Dirac}.  First one must replace the canonical variables $g_{ij}$ and
$\pi^{ij}$ by operators $\hat g_{ij}$ and $\hat \pi^{ij}=
-i\frac{\delta}{\delta g_{ij}}$ satisfying the canonical commutation
relations.\footnote{We choose units in which $\hbar$ and $c$ are $1$.}  One
then formally inserts these into the Hamiltonian functions $H(g,\pi)$ and
$H_i(g,\pi)$ of the classical theory to obtain operators $\widehat H(\hat
g,\hat\pi)$ and  $\widehat H_i(\hat g,\hat\pi)$ acting on functionals $\Psi(g)$
on the configuration space of 3-metrics.  Since the Hamiltonians were
constrained in the classical theory one demands that the corresponding
operators annihilate the physical states in the corresponding quantum
theory:
\begin{eqnarray} \widehat H \Psi & = & 0 \label{H0} \\
\widehat H_i \Psi & = & 0 \, . \label{Hi0}
\end{eqnarray}
Equation (\ref{Hi0}) has a simple meaning, namely that $\Psi(g)$ be
invariant under 3-diffeomorphisms (coordinate changes on the 3-manifold),
so that it depends on the 3-metric $g$ only through the
3-geometry. However, the interpretation of the {\it Wheeler-DeWitt equation}
(\ref{H0}) is not at all clear.

Before discussing the several problems which arise in attempts to give a
physical meaning to the approach just described, a few remarks are in
order: While we have omitted many technical details and problems from our
schematic description of the ``Dirac constraint quantization'' of gravity,
these problems either do not concern, or are consequences of, the main
conceptual problems of canonical quantum gravity. Other approaches, such  as
the canonical quantization of the Ashtekar formulation of classical general
relativity and its further development into loop quantum gravity, resolve some
of the technical problems faced by canonical quantization in the metric
representation, but leave the main conceptual problems  untouched.

Suppose now that we have found a solution $\Psi(g)$ to equations (\ref{H0})
and (\ref{Hi0}). What physical predictions would be implied? In orthodox
quantum theory a solution $\Psi_t$ of the time-dependent Schr\"odinger
equation provides us with a time-dependent probability distribution
$|\Psi_t|^2$, as well as with the absolute square of other time-dependent
probability amplitudes. The measurement problem and the like aside, the
physical meaning of these is reasonably clear: they are probabilities for the
results of the measurement of the configuration or of other
observables. But any attempt to interpret canonical quantum gravity along
orthodox lines immediately faces the following  problems:

\begin{itemize}
\item {\bf The problem of time}: In canonical quantum gravity there is no
time-dependent Schr\"odinger equation; it was replaced by the
time-independent Wheeler-DeWitt equation. The Hamiltonians---the generators
of multi-fingered-time evolution in the classical case---annihilate the
state vector and therefore cease to generate any evolution at all. The
theory provides us with only a timeless wave function on the configuration
space of 3-metrics, i.e., on the possible configurations of space, not of
space-time. But how can a theory that provides us (at best) with a single
fixed probability distribution for configurations of space ever be able to
describe the always changing world in which we live?  This, in a nutshell,
is the problem of time in canonical quantum gravity.
\item {\bf The problem of 4-diffeomorphism invariance}: The fundamental
symmetry at the heart of general relativity is its invariance under general
coordinate transformations of space-time. It is important to stress that
almost any theory can be formulated in such a 4-diffeomorphism invariant
manner by adding further structure to the theory (e.g., a preferred
foliation of space-time as a dynamical object). General relativity has what
is sometimes called serious diffeomorphism-invariance, meaning that it
involves no space-time structure beyond the 4-metric and, in particular,
singles out no special foliation of space-time. In canonical quantum
gravity, while the invariance under coordinate transformations of space is
retained, it is not at all clear what 4-diffeomorphism invariance could
possibly mean. Therefore the basic symmetry, and arguably the essence, of
general relativity seems to be lost in quantization.
\item {\bf The problem of ``no outside observer'':} One of the most
fascinating applications of quantum gravity is to quantum
cosmology. Orthodox quantum theory attains physical meaning only via its
predictions about the statistics of outcomes of measurements of
observables, performed by observers that are not part of the system under
consideration, and seems to make no clear physical statements about the
behavior of a closed system, not under observation. The quantum formalism
concerns the interplay between---and requires for its very meaning---two
kinds of objects: a quantum system and a more or less classical apparatus.
It is hardly imaginable how one could make any sense out of this formalism
for quantum cosmology, for which the system of interest is the whole
universe, a closed system if there ever was one.
\item {\bf The problem of diffeomorphism invariant observables:} Even if we
pretend for the moment that we are able to give meaning to the quantum
formalism without referring to an observer located outside of the universe,
we encounter a more subtle difficulty. Classical general relativity is
fundamentally diffeomorphism invariant. It is only the space-time geometry,
not the 4-metric nor the identity of the individual points in the
space-time manifold, that has physical significance.  Therefore the
physical observables in general relativity should be independent of special
coordinate systems; they must be invariant under 4-diffeomorphisms, which
are in effect generated by the Hamiltonians $H$ and $H_i$.  Since the
quantum observables are constructed, via  quantization, from the classical
ones, it would seem that  they must  commute with the Hamiltonians
$\widehat H$ and $\widehat H_i$. But such diffeomorphism invariant quantum
observables are extremely hard to come by, and there are certainly far too
few of them to even begin to account for the bewildering variety of our
experience which it is the purpose of quantum theory to explain.   (For a
discussion of the question of existence of diffeomorphism invariant
observables, see Kucha\v{r} \cite{KucharPoT}.)
\end{itemize}
These conceptual problems, and the attempts to solve them, have lead to a
variety of technical problems that are discussed in much detail in, e.g.,
Kucha\v{r} \cite{KucharPoT}, \cite{KucharCQG} and Isham \cite{Isham}. 
However, since we are
not aware of any orthodox proposals successfully resolving the conceptual
problems, we shall not discuss such details here. Rather, we shall proceed
in the opposite direction, toward their common cause, and argue that they
originate in a deficiency shared with, and inherited from, orthodox quantum
mechanics: the lack of a coherent ontology.

Regarding the first two problems of canonical quantum gravity, it is not
hard to discern their origin: the theory is concerned only with
configurations of and on space, the notion of a space-time having entirely
disappeared. It is true that even with classical general relativity,
Newton's external absolute time is abandoned. But a notion of time, for an
observer located somewhere in space-time and employing a coordinate system
of his convenience, is retained, emerging from space-time. The problem of time
in canonical quantum gravity is a direct consequence of the fact that in an
orthodox quantum theory for space-time itself we must insist on its
nonexistence (compare the quote at the beginning of this
article). Similarly, the problem of diffeomorphism invariance, or, better,
the problem of not even being able to address this question properly, is an
immediate consequence of having no notion of space-time in orthodox quantum
gravity.

\section{The basic problem of orthodox quantum theory: the lack of
a coherent ontology}

Despite its extraordinary predictive successes, quantum theory has,
since its inception some seventy-five years ago, been plagued by severe
conceptual difficulties. The most widely cited of these is the measurement
problem, best known as the paradox of Schr\"odinger's cat. For many
physicists the measurement problem is, in fact, not {\it a} but {\it the}
conceptual difficulty of quantum theory.

In orthodox quantum theory the wave function of a physical system is
regarded as providing its complete description. But when we analyze the
process of measurement itself in quantum mechanical terms, we find that the
after-measurement wave function for system and apparatus arising from
Schr\"odinger's equation for the composite system typically involves a
superposition over terms corresponding to what we would like to regard as
the various possible results of the measurement---e.g., different pointer
orientations. Since it seems rather important that the actual result of the
measurement be a part of the description of the after-measurement
situation, it is difficult to believe that the wave function alone provides
the complete description of that situation.

The usual collapse postulate for quantum measurement solves this problem
for all practical purposes, but only at the very steep price of the
introduction of an {\it observer\/} or {\it classical measurement
apparatus\/} as an irreducible, unanalyzable element of the theory.  This
leads to a variety of further problems. The unobserved physical reality
becomes drastically different from the observed, even on the macroscopic
level of daily life. Even worse, with the introduction at a fundamental
level of such vague notions as classical measurement apparatus, the
physical theory itself becomes unprofessionally vague and ill defined. The notions of
observation and measurement can hardly be captured in a manner appropriate
to the standards of rigor and clarity that should be demanded of a
fundamental physical theory. And in quantum cosmology the notion of an
external observer is of course entirely obscure.

The collapse postulate is, in effect, an unsuccessful attempt to evade the
measurement problem without taking seriously its obvious implication: that
the wave function does not provide a complete description of physical
reality. If we do accept this conclusion, we must naturally inquire about
the nature of the more complete description with which a less problematical
formulation of quantum theory should be concerned. We must ask, which
theoretical entities, in addition to the wave function, might the theory
describe? What mathematical objects and structures represent entities that,
according to the theory, simply {\it are\/}, regardless of whether or not
they are observed? We must ask, in other words, about the {\it primitive
ontology\/} of the theory, what the theory is fundamentally about (see
Goldstein \cite{Shelly}). And when we know what the theory is really about,
measurement and observation become secondary phenomenological concepts
that, like anything else in a world governed by the theory, can be analyzed
in terms of the behavior of its primitive ontology.

By far the simplest possibility for the primitive ontology is that of
particles described by their positions. The corresponding theory, for
non-rela\-tivistic particles, is Bohmian mechanics.

\section{Bohmian mechanics}

According to Bohmian mechanics  the complete description
of an $n$-particle system is provided by its wave function $\Psi$ 
together with its configuration $Q = ({\bf Q}_1,\ldots, {\bf Q}_n)$, where the 
${\bf Q}_k$ are the positions of its particles. The wave function,
which evolves according to Schr\"odinger's equation, choreographs the
motion of the particles: these evolve---in the simplest manner
possible---according to a first-order ordinary differential equation
\[
\frac{dQ}{dt}=v^\Psi(Q)
\]
whose right hand side, a velocity vector field on configuration space, is
generated by the wave function. Considerations of simplicity and space-time
symmetry---Galilean and time-reversal invariance---then determine the form
of $v^\Psi$, yielding the defining (evolution) equations of Bohmian
mechanics (for spinless particles):
\begin{equation} \label{Bohm}
\frac{d{\bf Q}_k}{dt} = {\bf v}^\Psi_k({\bf Q}_1,\ldots, {\bf Q}_n) =
\frac{ \hbar}{m_k}\mbox{Im}\frac{\nabla_{{\bf q}_k}\Psi}{\Psi}({\bf
Q}_1,\ldots, {\bf Q}_n) 
\end{equation}
and
\begin{equation} \label{Schroed}
i\hbar\frac{\partial\Psi}{\partial t}=\widehat H\Psi
\end{equation}
where $\widehat H$ is the usual Schr\"odinger Hamiltonian, containing as
parameters the masses $m_1,\ldots,m_n$ of the particles as well as the
potential energy function $V$ of the system. For an $n$-particle
universe, these two equations form a complete specification of the
theory. There is no need, and indeed no room, for any further axioms,
describing either the behavior of other ``observables'' or the effects
of ``measurement.''

Bohmian mechanics is the most naively obvious embedding imaginable of
Schr\"odinger's equation into a completely coherent physical theory!  If
one didn't already know better, one would naturally conclude that it can't
``work,'' i.e., that it can't account for quantum phenomena. After all, if
something so obvious and, indeed, so trivial works, great physicists---so
it would seem---would never have insisted, as they have and as they
continue to do, that quantum theory demands radical epistemological and
metaphysical innovations.

Moreover, it is hard to avoid wondering  how  Bohmian mechanics {\it could}
have much to do with quantum theory? Where is quantum randomness in
this deterministic theory? Where is quantum uncertainty? Where are 
operators as observables and all the rest?

Be that as it may, Bohmian mechanics is certainly {\it a} theory. It
describes a world in which particles participate in a highly non-Newtonian
motion, and it would do so even if this motion had absolutely nothing to do
with quantum mechanics.

It turns out, however, as a surprising consequence of the equations
(\ref{Bohm}) and (\ref{Schroed}), that when a system has wave function
$\Psi$, its configuration is typically random, with probability density
$\rho$ given by $\rho=|\Psi|^2$, the {\it quantum equilibrium} distribution. In
other words, it turns out that systems are somehow typically in quantum
equilibrium. Moreover, this conclusion comes together with the
clarification of what precisely this means, and also implies that a Bohmian
universe embodies an absolute uncertainty which can itself be regarded as
the origin of the uncertainty principle. We shall not go into these matters
here, since we have discussed them at length elsewhere (D\"urr, 
Goldstein and Zangh\`\i\ \cite{QE}). We note,
however, that nowadays, with chaos theory and nonlinear dynamics so
fashionable, it is not generally regarded as terribly astonishing for an
appearance of randomness to emerge from a deterministic dynamical system.

It also turns out that the entire quantum formalism, operators as
observables and all the rest, is a consequence of Bohmian mechanics,
emerging from an analysis of idealized measurement-like situations (for
details, see Daumer et al. \cite{Naive} and \cite{Operators}; see also Bohm \cite{Bohm52}).
There is no measurement problem in Bohmian mechanics because the complete
description of the after-measurement situation includes, in addition to the
wave function, the definite configuration of the system and
apparatus. While the wave function may still be a superposition of states
corresponding to macroscopically different possible outcomes, the actual
configuration singles out the outcome that has occurred.

Why have we elaborated in such detail on non-relativistic quantum mechanics
and Bohmian mechanics if our main concern here is with quantum gravity?
Because there are two important lessons to be learned from a Bohmian
perspective on quantum theory. First of all, the existence of Bohmian
mechanics demonstrates that the characteristic features of quantum theory,
usually viewed as fundamental---intrinsic randomness, operators as
observables, non-commutativity, and uncertainty---need play no role
whatsoever in the formulation of a quantum theory, naturally emerging
instead, as a consequence of the theory, in special measurement-like
situations. Therefore we should perhaps not be too surprised when
approaches to quantum gravity that regard these features as fundamental
encounter fundamental conceptual difficulties.  Second, the main point of
our paper is made transparent in the simple example of Bohmian
mechanics. If we base our theory on a coherent ontology, the conceptual
problems may disappear, and, what may be even more important, a genuine
understanding of the features that have seemed most puzzling might be
achieved.

We shall now turn to what one might call a Bohmian approach to quantum
gravity.

\section{Bohmian quantum gravity}

The transition from quantum mechanics to Bohmian mechanics is very simple,
if not trivial: one simply incorporates the actual configuration into the
theory as the basic variable, and stipulates that this evolve in a natural
way, suggested by symmetry and by Schr\"odinger's equation.  The velocity
field $v^{\Psi_t}$ is, in fact, related to the quantum probability current
$j^{\Psi_t}$ by
\[ 
v^{\Psi_t} = \frac{j^{\Psi_t}}{|\Psi_t|^2}\, ,
\]
suggesting, since $\rho^\Psi =|\Psi|^2$ satisfies the continuity equation
with $j^{\Psi_t} = \rho^{\Psi_t} v^{\Psi_t}$, that the empirical
predictions of Bohmian mechanics, for positions and ultimately, in fact,
for other ``observables'' as well, agree with those of quantum mechanics (as
in fact they do; see D\"urr et al. \cite{QE}).

Formally, one can follow exactly the same procedure in canonical quantum
gravity, where the configuration space is the space of (positive-definite)
3-metrics (on an appropriate fixed manifold). The basic variable in Bohmian
quantum gravity is
therefore the 3-metric $g$ (representing the geometry on a
space-like hypersurface of the space-time to be generated by the
dynamics)  and its change under (what will become)
normal deformations is given by a vector field on configuration space
generated by the wave function $\Psi(g)$. Considerations analogous to those for
 non-relativistic particles lead to the following form for the
Bohmian equation of motion:
\begin{equation} \label{GBohm}
d g_{ij}(x^a)= G_{ijab}(x^a)\mbox{Im}\left( \Psi(g)^{-1}\frac{\delta
\Psi(g)}{\delta g_{ab}(x^a)}\right) N(x^a)\,d\tau\, .
\end{equation} 
The wave function $\Psi(g)$ is a solution of the timeless Wheeler-DeWitt
equation (\ref{H0}) and therefore does not evolve. But the vector field on
the right hand side of (\ref{GBohm}) that it generates is typically
nonvanishing if $\Psi(g)$ is complex, leading to a nontrivial evolution
$g(\tau)$ of the 3-metric. Suitably gluing together the 3-metrics
$g(\tau)$, we obtain a space-time (see the paragraph after equation
(\ref{Hamil})).  Interpretations of canonical quantum gravity along these
lines have been proposed by, e.g., Holland \cite{Holland} and discussed, e.g., by
Shtanov  \cite{Shtanov}. Minisuperspace Bohmian cosmologies have been considered by Kowalski-Glikman and Vink \cite{Kowalski}, Squires \cite{Squires} and 
Callender and Weingard \cite{Callender}.

However, there is a crucial point which is often overlooked or, at least,
not made sufficiently clear in the literature. A space-time generated by a
solution of (\ref{H0}) via (\ref{GBohm}) will in general depend on the
choice of lapse function $N$ (or $N(\tau)$). Thus the theory is not well
defined as so far formulated. There are essentially two ways to complete
the theory. Either one chooses a special lapse function $N$, e.g., $N=1$,
or one employs only special solutions $\Psi$ of (\ref{H0}), those yielding
a vector field that generates an $N$-independent space-time.  In the first
case, with special $N$ but general solution $\Psi$ of (\ref{H0}), the
general covariance of the theory will typically be broken, the theory
incorporating a special foliation (see the paragraph before the one
containing equation (\ref{H0})).  The possible existence of special
solutions giving rise to a covariant dynamics will be discussed in more
detail elsewhere (Goldstein and Teufel, \cite{ShellyStef}), and will be touched upon towards the
end of Section 6.  However, most the following discussion, especially in
the first part of Section 6, does not depend upon whether or not the theory
incorporates a special foliation.

Let us now examine the impact of the Bohmian formulation of canonical
quantum gravity on the basic conceptual problems of orthodox canonical
quantum gravity.  Since a solution to the equations of Bohmian quantum
gravity defines a space-time, the problem of time is resolved in the most
satisfactory way: Time plays exactly the same role as in classical general
relativity; there is no need whatsoever for an external absolute time,
which has seemed so essential to orthodox quantum theory.  The problem of
diffeomorphism-invariance is ameliorated, in that in this formulation it is
at least clear what diffeomorphism-invariance means. But, as explained
above, general covariance can be expected at most for special solutions of
(\ref{H0}). If it should turn out, however, that we must abandon general
covariance on the fundamental level by introducing a special foliation of
space-time, it may still be possible to retain it on the observational
level (see, e.g., M\"unch-Berndl et al. \cite{BDGZ}, where it is also
argued, however, that a special, dynamical, foliation of space-time need
not be regarded as incompatible with serious covariance).

A short answer to the problems connected with the role of observers and
observables is this: There can be no such problems in the Bohmian
formulation of canonical quantum gravity since observers and observables
play no role in this formulation. But this is perhaps too short. What,
after all, is wrong with the observation that, since individual space-time
points have no physical meaning, physically significant quantities must
correspond to diffeomorphism-invariant observables, of which there are far
too few to describe very much of what we most care about?  

	The basic answer, we believe, is this: We ourselves are not---or,
at least, need not be\footnote{In some models of quantum cosmology, e.g.,
those permitting the definition of a global time function, it may well be
possible to pick ourselves out in a diffeomorphism-invariant
manner.}---diffeomorphism invariant: Most physical questions of relevance
to us are not formulated in a diffeomorphism invariant manner because,
naturally enough, they refer to our own special location in space-time.
Nonetheless, we know very well what they mean---we know, e.g., what it
means to ask where and when something happens with respect to our own point
of view. Such questions can be addressed, in fact because of diffeomorphism
invariance, by taking into account the details of our environment and
asking about the local predictions of the theory conditioned on such an
environment, past and present.

	The observer who sets the frame of reference for his physical
predictions is part of and located inside the system---the universe.  In
classical general relativity this is not at all problematical, since that
theory provides us with a coherent ontology, a potentially complete
description of space-time and, if we wish, a description taking into
account our special point of view in the universe. But once the step to
quantum theory is taken, the coherent space-time ontology is replaced by an
incoherent ``ontology'' of quantum observables. In orthodox quantum theory
this problem can be talked away by introducing an outside observer actually
serving two purposes: the observer sets the frame of reference with respect
to which the predictions are to be understood, a totally legitimate and
sensible purpose. But of course the main reason for the focus on observers
in quantum theory is that it is only with respect to them that the
intrinsically incoherent quantum description of the system under
observation can be given any meaning.  In quantum cosmology, however, no
outside observer is at hand, neither for setting a frame of reference nor
for transforming the incoherent quantum picture into a coherent one.

	In Bohmian quantum gravity, again, both problems disappear. Since
we have a coherent description of the system itself, in this case the
universe, there is no need for an outside observer in order to give meaning
to the theory. Nor do we have to worry about the diffeomorphism invariance
of observables, since we are free to refer to observers who are themselves
part of the system.

	There is, however, an important aspect of the problem of time that
we have not yet addressed. From a Bohmian perspective, as we have seen, a
time-dependent wave function, satisfying Schr\"odinger's equation, is by no
means necessary to understand the possibility of what we call
change. Nonetheless, a great deal of physics is, in fact, described by such
time-dependent wave functions. We shall see in the next section how these
also naturally emerge from the structure of Bohmian quantum gravity, which
fundamentally has only a timeless universal wave function.

\section{A universal Bohmian theory}

When Bohmian mechanics is viewed from a universal perspective, the status
of the wave function is dramatically altered. To appreciate what we have in
mind here, it might help to consider  two very common objections to
Bohmian mechanics. 

Bohmian mechanics violates the action-reaction principle that is central to
all of modern physics, both classical and (non-Bohmian) quantum: In Bohmian
mechanics there is no back-action of the configuration upon the wave
function, which evolves, autonomously, according to Schr\"odinger's
equation. And the wave function, which is part of the state description
of---and hence presumably part of the reality comprising---a Bohmian
universe, is not the usual sort of physical field on physical space (like
the electromagnetic field) to which we are accustomed, but rather a field
on the abstract space of all possible configurations, a space of enormous
dimension, a space constructed, it would seem, by physicists as a matter of
convenience.

It should be clear by now what, from a universal viewpoint, the answer to
these objections must be: As first suggested by D\"urr et al.\,\cite{Meaning}, 
the wave
function $\Psi$ of the universe should be regarded as a representation, not
of substantial physical reality, but of physical law.  In a universal
Bohmian theory $\Psi$ should be a functional of the configurations of all
elements of physical reality: geometry, particle positions, field or string
configurations, or whatever primitive ontology turns out to describe nature
best. As in the case of pure quantum gravity, $\Psi$ should be a (special)
solution of some fundamental equation (such as the Wheeler-DeWitt equation
(\ref{H0}) with additional terms for particles, fields, etc.). Such a
universal wave function would be static---a wave function whose
timelessness constitutes the problem of time in canonical quantum
gravity---and, insofar as our universe is concerned, unique. But this
doesn't mean, as we have already seen, that the world it describes would be
static and timeless. No longer part of the state description, the universal
wave function $\Psi$ provides a representation of dynamical law, via the
vector field on configuration space that it defines. As such, the wave
function plays a role analogous to that of the Hamiltonian function
$H=H(Q,P)\equiv H(\xi)$ in classical mechanics---a function on phase space,
a space even more abstract than configuration space. In fact, the wave
function and the Hamiltonian function generate motions in pretty much the
same way
\[ \frac{d\xi}{dt}= \mbox{Der} H  
 \longleftrightarrow
\frac{dQ}{dt} = \mbox{Der}(\log \Psi)\, ,
\]
with Der a derivation. And few would be tempted to regard the Hamiltonian
function $H$ as a real physical field, or expect any back-action of
particle configurations on this Hamiltonian function.

Once we recognize that the role of the wave function is thus nomological,
two important questions naturally arise: Why and how should a formalism
involving time-dependent wave functions obeying Schr\"odinger's equation
emerge from a theory involving a fixed timeless universal wave function?
And which principle singles out the special unique wave function $\Psi$
that governs the motion in our universe?  Our answers to these questions
are somewhat speculative. But they do provide further insight into the role
of the wave function in quantum mechanics and might even explain why, in
fact, our world is quantum mechanical.

In order to understand the emergence of a time-dependent wave function, we
must ask the right question, which is this: Is it ever possible to find a
simple effective theory governing the behavior of suitable subsystems of a
Bohmian universe?  Suppose, then, that the configuration of the universe
has a decomposition of the form $q=(x,y)$, where $x$ describes the degrees
of freedom with which we are somehow most directly concerned (defining the
{\it subsystem\/}, the ``$x$-system'') and $y$ describes the remaining
degrees of freedom (the subsystem's {\it environment\/}, the
``$y$-system''). For example, $x$ might be the configuration of all the
degrees of freedom governed by standard quantum field theory, describing
the fermionic matter fields as well as the bosonic force fields, while $y$
refers to the gravitational degrees of freedom.  Suppose further that we
have, corresponding to this decomposition, a solution
$Q(\tau)=(X(\tau),Y(\tau))$ of the appropriate (yet to be defined)
extension of (\ref{GBohm}), where the real continuous parameter $\tau$
labels the slices in a suitable foliation of space-time.

Focus now on the {\it conditional wave function\/}
\begin{equation}
\psi_\tau(x) = \Psi(x,Y(\tau))
\end{equation}
of the subsystem, governing its motion, and ask whether $\psi_\tau(x)$
could be---and might under suitable conditions be expected to be---governed
by a simple law that does not refer directly to its environment. (The
conditional wave function of the $x$-system should be regarded as defined
only up to a factor that does not depend upon $x$.)

Suppose that $\Psi$ satisfies an equation of the form (\ref{H0}), with
$\widehat H=\{{\widehat H}_N\}$. Suppose further that for $y$ in some
``$y$-region'' of configuration space and for some choice of lapse function
$N$ we have that ${\widehat H}_N\simeq {\widehat H}_N^{(x)}+ {\widehat
H}_N^{(y)}$ and can write
\begin{eqnarray}
\Psi(x,y) & = &\lefteqn{e^{-i\tau \widehat H_N}\Psi(x,y) \simeq e^{-i\tau \widehat H_N}\sum_\alpha
\psi^\alpha_0(x) \phi^\alpha_0 (y)} \nonumber\\
&  \simeq & \sum_\alpha \left(
e^{-i\tau \widehat H_N^{(x)}} \psi^\alpha_0(x) \right)\left( e^{-i\tau \widehat H_N^{(y)}}
\phi^\alpha_0 (y) \right) \nonumber\\ & =: & \sum_\alpha  \psi^\alpha_\tau(x)
\phi^\alpha_\tau (y) \label{CL}
\end{eqnarray}
where the $\phi^\alpha_0$ are ``narrow disjoint wave packets'' and remain
approximately so as long as $\tau$ is not too large. Suppose (as would be
the case for Bohmian mechanics) that the motion is such that if the
configuration $Y(0)$ lies in the support of one
$\phi^{\alpha'}_0$, then $Y(\tau)$ will keep up with $\phi^{\alpha'}_\tau$ as
long as the above conditions are satisfied. It then follows from
(\ref{CL}) that for the conditional wave function of the subsystem we have
\[\psi_\tau(x) \approx \psi^{\alpha'}_\tau\, ,
\] 
and it thus approximately satisfies the time-dependent Schr\"odinger
equation
\begin{equation}\label{se}  
i\frac{\partial \psi}{\partial \tau} = \widehat H_N^{(x)} \psi \, .
\end{equation}  
(In the case of (an extension of) Bohmian quantum gravity with preferred
foliation, this foliation must correspond to the lapse function $N$ in
(\ref{CL}).)

We may allow here for an interaction $\widehat W_N(x,y)$ between the
subsystem and its environment in the Hamiltonian in (\ref{CL}), provided
that the influence of the $x$-system on the $y$-system is negligible. In
this case we can replace $\widehat H_N^{(x)}$ in (\ref{CL}) and (\ref{se})
by $\widehat H_N^{(x)}(Y(\tau))\equiv \widehat H_N^{(x)}+\widehat
W_N(x,Y(\tau))$, since the wave packets $\phi^\alpha(y)$ are assumed to be
narrow. Think, for the simplest example, of the case in which the
$y$-system is the gravitational field and the $x$-system consists of very
light particles.

Now one physical situation (which can be regarded as corresponding to a
region of configuration space) in which (\ref{CL}), and hence the
Schr\"odinger evolution (\ref{se}), should obtain is when the $y$-system
behaves semiclassically: In the semiclassical regime, one expects an
initial collection of narrow and approximately disjoint wave packets
$\phi^\alpha_0(y)$ to remain so under their (approximately classical)
evolution.

As a matter of fact, the emergence of Schr\"odinger's equation in the
semiclassical regime for gravity can be justified in a more systematic way,
using perturbation theory, by expanding $\Psi$ in powers of the
gravitational constant $\kappa$.  Then for a ``semiclassical wave
function'' $\Psi$, the phase $S$ of $\Psi$, to leading order, $\kappa^{-1}$,
depends only on the 3-metric and obeys the classical
Einstein-Hamilton-Jacobi equation, so that the metric evolves approximately
classically, with the conditional wave function for the matter degrees of
freedom satisfying, to leading (zeroth) order, Schr\"odinger's equation
for, say, quantum field theory on a given evolving background. The relevant
analysis was done by Banks \cite{Banks} for canonical quantum gravity, but
the significance of that analysis is rather obscure from an orthodox
perspective: 

The semiclassical limit has been proposed as a solution to the problem of
time in quantum gravity, and as such has been severely criticized by Kucha\v{r}
\cite{KucharPoT}, who concludes his critique by observing that ``the
semiclassical interpretation does not solve the standard problems of
time. It merely obscures them by the approximation procedure and, along the
way, creates more problems.''  Perhaps the main difficulty is that, within
the orthodox framework, the classical evolution of the metric is not really
an approximation at all. Rather, it is put in by hand, and can in no way be
justified on the basis of an entirely quantum mechanical treatment, even as
an approximation. This is in stark contrast with the status of the
semiclassical approximation within a Bohmian framework, for which there is
no problem of time. In this approach, the classical evolution of the metric
is indeed merely an approximation to its exact evolution, corresponding to
the exact phase of the wave function (i.e., to equation (\ref{GBohm})). To
the extent that this approximation is valid, the appropriate conclusions
can be drawn, but the theory makes sense, and suffers from no conceptual
problems, even when the approximation is not valid.

Now to our second question. Suppose that we demand of a universal dynamics
that it be first-order for the variables describing the primitive ontology
(the simplest possibility for a dynamics) and covariant---involving no
preferred foliation, no special choice of lapse function $N$, in its
formulation. This places a very strong constraint on the vector field
defining the law of motion---and on the universal wave function, should this
motion be generated by a wave function. The set of wave functions
satisfying this constraint should be very small, far smaller than the set
of wave functions we normally consider as possible initial states for a
quantum system.  However, according to our conception of the wave function
as nomological, this very fact might well be a distinct virtue.

We have begun to investigate the possibility of a first-order covariant
geometrodynamics in connection with Bohmian quantum gravity, and have found
that the constraint for general covariance is captured by the Dirac algebra
(see also Hojman, Kucha\v{r} and Teitelboim \cite{Hojman}), which expresses
the relation between successive infinitesimal deformations of
hypersurfaces, taken in different orders: We have shown (see Goldstein and
Teufel \cite{ShellyStef}) that defining a representation of the Dirac
algebra is more or less the necessary and sufficient condition for a vector
field on the space of 3-metrics to yield a generally covariant dynamics,
generating a 4-geometry involving no dynamically distinguished
hypersurfaces.

This work is very much in its infancy. In addition to the problem of
finding a mathematically rigorous proof of the result just mentioned, there
remains the difficult question of the possible representations of the Dirac
algebra, both for pure gravity and for gravity plus matter. For pure
gravity it seems that a first-order generally covariant geometrodynamics is
achievable, but only with vector fields that generate classical
4-geometries---solutions of the Einstein equations with a possible
cosmological constant. How this situation might be affected by the
inclusion of matter is not easy to say. 

Even a negative result---to the effect that a generally covariant Bohmian
theory must involve additional space-time structure---would be
illuminating.  A positive result---to the effect that a first-order
dynamics, for geometry plus matter, that does not invoke additional
space-time structure can be generally covariant more or less only when the
vector field defining this dynamics arises from an essentially unique wave
function of the universe that happens to satisfy an equation like the
Wheeler-DeWitt equation (and from which a time-dependent Schr\"odinger
equation emerges, in the manner we've described, as part of a
phenomenological description governing the behavior of appropriate
subsystems)---would be profound. For then we would know, not just what
quantum mechanics is, but why it is.

\end{document}